\documentclass[doublecol]{epl2}

\usepackage{graphicx}
\usepackage{epsfig}
\usepackage{subfigure}
\usepackage{sidecap}
\usepackage{color}
\usepackage{epsf}

\title{Effect of an Impurity on Grey Soliton Dynamics in Cigar-Shaped Bose-Einstein Condensate}

\author{Priyam Das \inst{1}\thanks{\email{priyam@iiserkol.ac.in}} \and Sumona Gangopadhyay \inst{2} \and Prasanta K. Panigrahi \inst{1}}
\shortauthor{Priyam Das \etal}
	
 \institute{
	     \inst{1} Indian Institute of Science Education and Research - Kolkata, Mohanpur, Nadia 741252, India \\
	     \inst{2} Indian Association for the Cultivation of Science, Jadabpur, Kolkata 700032, India
           }
 

\abstract{
In a cigar shaped Bose-Einstein condensate, explicit solutions of
the coupled mean-field equations, describing defect-grey soliton
dynamics are obtained, demonstrating the coexistence of grey soliton
and a localized defect.	 Unlike the case of dark soliton, where the
defect trapping center has vanishing superfluid density, the moving
grey soliton necessarily possesses a finite superfluid component at
the defect location. The wave vector of the impurity is controlled
by the velocity of the grey soliton, which has an upper bound. It is
found that the presence of the impurity lowers the speed of the grey
soliton, as compared to the defect free case, where it can reach the
sound velocity. The grey soliton's energy gets substantially
modified through its interaction with the defect, opening up the
possibility of its control through defect dynamics.
}

\begin{document}

\maketitle

Quite some time back, Gross formulated the mean field equations
describing the macroscopic dynamics of defects and superfluid matter
\cite{Gross1,Gross2}. Akin to the Gross-Pitaevskii (GP) equation for
Bose-Einstein condensate (BEC), these mean-field equations are well
suited for describing the defect-BEC dynamics in a trap. On the
theoretical side, a number of works explored the phenomenological
features of BEC-impurity complex
\cite{Gold,Chin,Capuzzi,Giacconi,Cavalcanti}. The dynamics of the BEC,
trapped in an optical lattice, in presence of a localized impurity has
been investigated \cite{Brazhnyi1}. The interactions of defect atoms in
an optical lattice with a uniform BEC have also been studied for the
understanding the dephasing effect on this system \cite{Jaksch}. Recently,
Roberts and Rica \cite{Roberts}, investigated the behavior of the
impurity field in a BEC and identified the parameter domains for the
formation of a crystal of impurity fields and supersolid phases. The
fact that the geometry of this complex can be effectively manipulated,
has motivated the study of the BEC-defect dynamics in lower
dimensions. Cigar shaped BEC has received special attention, because
of the possibility of identifying exact solutions. It was found that,
even approximate solutions in one dimension, well represent this
coupled system \cite{Brudered}.

Experimental realization of dark and bright soliton
\cite{Burger,Denschlag,Khaykovich,Eiermann,Cornish,Becker}, and soliton trains
\cite{Strecker} has given impetus to the study of the interaction
between defect-soliton system. In $1997$, Konotop \etal
\cite{Konotop}, studied the dark soliton-impurity complex by means of
a modified adiabatic approximation and observed differences between
dark and bright soliton dynamics perturbed by a point defect. In
Ref. \cite{Kivshar}, Frantzeskakis \etal, considered a static
impurity and investigated the interaction of dark soliton with
localized impurities. It was found that the dark soliton can get
reflected or transmitted by a repulsive impurity. Self-trapping of
impurities was studied, for the case of both repulsive and attractive
interactions \cite{Brudered}. The soliton-defect dynamics was also
investigated by Goodman \etal  \cite{Goodman}. In a recent
work, it has been shown that a dissipation source can be used to
generate dark, bright, gap and ring dark solitons, by controlling
their phase and amplitude \cite{Brazhnyi2}. Very recently, Dries \etal \cite{Dries}, investigated the effects of impurities on collective dipole motion of the BEC 
and characterized the breakdown of superfluidity of the trapped cloud in both 3D
Thomas-Fermi and quasi-1D weakly interacting regimes. Apart from dark and bright
solitons, GP equation supports a complex envelope soliton, analogous
to the Bloch soliton in magnetic systems \cite{Lieb,Kulish}. This grey
soliton has recently been observed through a controlled density
engineering method \cite{Weller,Shomroni}.  Its velocity can take
values from zero to the sound velocity and has a dispersion very
different from that of dark soliton \cite{Komineas,Jackson1}. It is
then natural to inquire the nature of the interaction between grey
soliton and a point like defect.

In a recent experiment, it was observed that, when the velocity of
the impurity atom was made to decrease below the condensate sound
velocity, the collisional cross section decreased abruptly
\cite{Ketterle}. This indicates the better applicability of
mean-field dynamics in this regime. Keeping this in mind, we study
here the BEC-defect complex in the mean-field approximation, in a
quasi-one dimensional scenario, where the defect atom moves in the
condensate with constant velocity. As will be seen below, for this
system both defect and soliton velocities are bounded from above.
The general case is investigated, where defect and the condensate
atoms have different masses. The density of the impurity atom is
assumed to be small, so that the effect of the impurity on BEC
excitation spectrum remains negligible. We find exact solutions of
the coupled mean-field equations, describing the interaction of a
grey soliton with a localized defect. The local minimum of the
soliton in the defect location has a non-vanishing superfluid
density. The presence of the defect lowers the grey soliton's velocity, 
as compared to the defect free case, where the maximum velocity is the 
sound velocity. Physically, the presence of the defect atom acts like 
a drag on the grey soliton, which results in a decrease of its speed 
from the sound velocity. The stability of the above solution is 
investigated using the well known Vakhitov-Kolokolov (VK) criteria
\cite{Vakhitov,Weinstein} and found that the obtained solutions are
stable. It is also found that the defect affects the energy of the
grey soliton considerably, opening up the possibility of controlling
the grey soliton through defect dynamics. We then compared these
results with pure dark and bright soliton-defect complex, obtained
from the mean-field equations. The mean-field solutions require the
masses of the impurity and that of the condensate atoms to be same,
implying that the impurity is the same atom in a different hyperfine
level. Unlike the previous case, no restriction is found on the
velocity of the soliton or that of the defect.

In the model of Gross, the interaction between impurity and
the condensate is treated in the Hartree approximation.
The defect-condensate dynamics is described by
the following coupled equations, where the wavefunctions of the
condensate atoms and the defect are $\psi_{a}$ and $\psi_{b}$,
respectively:

\begin{eqnarray}
i \hbar \frac{\partial \Psi_{a}}{\partial t} &=&
-\frac{\hbar^{2}}{2m_{a}} \vec{\nabla^{2}}\Psi_{a} + V_{ext}
\Psi_{a} + g |\Psi_{a}|^{2} \Psi_{a} \nonumber \\& & \hskip1.75cm +
\kappa
|\Psi_{b}|^{2} \Psi_{a} - \mu \Psi_{a}, \label{atom_eq1} \\
\textrm{and \,\,\,} i\hbar\frac{\partial
  {\Psi_b}}{\partial t} &=& - \frac{\hbar^2}{2m_b}
\vec{\nabla^2}\Psi_b + V_{ext} \Psi_{b} + \kappa |\Psi_a|^2\Psi_b.
\label{def_eq2}
\end{eqnarray}
Here, $g = \frac{4 \pi \hbar^{2} a}{m_{a}}$ is the strength of the
atom-atom interaction and the strength of the interaction between
atoms in the condensate and the impurity is given by, $\kappa =
\frac{2 \pi \hbar^{2} a_{12}}{m_{r}}$, where, $m_{r} =
\frac{m_{a}m_{b}}{m_{a} + m_{b}}$ is the reduced mass. The mass of
the condensate atoms and that of the defect are $m_{a}$ and $m_{b}$,
respectively, with $\mu$ being the chemical potential. The
condensate wavefunction is normalized to the number of atoms $N$ and
the defect wavefunction to unity.

For cigar shaped BEC, the three dimensional coupled GP equation can be
transformed into quasi-one dimension: $\Psi_{i}(r,t) = \psi_{i}(x,t)
f_{i}(y,z)$, ($i = a,b$). The condensate atoms and the impurity can be
made to experience the same trapping potential $V_{ext} = \frac{1}{2}
m_{a} \omega^{2}_{a} (y^{2} + z^{2})$, by tuning the axial frequencies
of the atoms $\omega_{a}$ and that of the impurity $\omega_{b}$, which
differs by a factor of $\sqrt{m_{a}/m_{b}}$ \cite{Modugno}. As is
known, two different masses can bring in transverse separation due to
gravity \cite{Modugno,Weiman}. We assume that, this separation is
negligible. In the reduction to quasi-one dimension, taking into
account the tight harmonic trap, one needs two different widths for
the transverse Gaussian profiles
\cite{Salasnich1,Salasnich2,Salasnich3}:
\begin{eqnarray}
  f_{a}(y,z) &=& \frac{1}{\sqrt{\pi} a_{\perp}} e^{- (y^{2} + z^{2})/2 a^{2}_{\perp}}, \\ \textrm{\,\,\,and\,\,\,\,\,\,}
f_{b}(y,z) &=& \frac{1}{\sqrt{\pi} b_{\perp}} e^{- (y^{2} + z^{2})/2
b^{2}_{\perp}},
\end{eqnarray}
where $a_{\perp} = \sqrt{\frac{\hbar}{m_{a} \omega_{a}}}$ and
$b_{\perp} = \sqrt{\frac{\hbar}{m_{b} \omega_{b}}}$. The three
dimensional coupled equations can be mapped into one dimension by
minimizing the action functional after integrating over the transverse
degrees of freedom. In the weak coupling scenario, the coupled
equations in one dimension are given by
\cite{Jackson2,Salasnich4},
\begin{eqnarray}
i \hbar \frac{\partial \psi_{a}}{\partial t} &=& -\frac{\hbar^{2}}{2
  m_{a}} \frac{\partial^{2} \psi_{a}}{\partial x^{2}} + \tilde{g} |\psi_{a}|^{2}
\psi_{a} + \tilde{\kappa} |\psi_{b}|^2 \psi_{a} - \mu
\psi_{a}, \label{GP1d_a}\nonumber\\ \\ i \hbar \frac{\partial
  \psi_{b}}{\partial t} &=& -\frac{\hbar^{2}}{2 m_{b}} \frac{\partial^{2}
  \psi_{b}}{\partial x^{2}} + \tilde{\kappa} |\psi_{a}|^2
\psi_{b}, \label{GP1d_b}
\end{eqnarray}
where $\tilde{g} = \frac{1}{2 \pi a^{2}_{\perp}} g$ and
$\tilde{\kappa} = \frac{1}{\pi(a^{2}_{\perp} + b^{2}_{\perp})}
\kappa$. We now drop the tilde for notational convenience and study
the general case, where the mass of the condensate atoms and that of the
impurity are different. The following envelope profiles lead to the
description of grey soliton impurity complex: $\psi_{a} =
\sqrt{\sigma_{a}}e^{i \chi}$ and $\psi_{b} = \sqrt{\sigma_{b}}e^{i k x
  - i \omega t}$. It is worth emphasizing that for the grey soliton to
exist, the defect should necessarily possess the plain wave component
$e^{i k x}$. Current conservation and consistency conditions yield:
\begin{equation}
\frac{\hbar}{m_{a}} \chi' =  u (1 -
\frac{\sigma_{0}}{\sigma_{a}}) \label{phase} \textrm{\,\,\,\,and\,\,\,\,}
\hbar k = m_{b} u.
\end{equation}

\begin{figure}[t]
\begin{center}
\includegraphics[scale = 0.3]{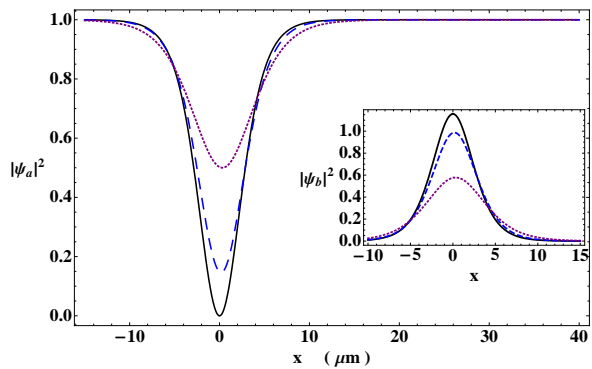}
\caption{The figure shows the variation of the grey soliton profile
  for different values of $\theta$. The solid (black), dashed (blue)
  and dotted (purple) lines correspond to $\theta = 0$, $\pi/8$ and
  $\pi/4$, respectively. Inset depicts the corresponding density
  profiles of the impurity atom. The densities ($|\psi_{i}|^{2}, i = a,b$) are measured in the unit
  of $\sigma_{0}$. The parameter values are: $\sigma_{0} a = 0.1$, $\omega_{a} = 2 \pi \times 120$ Hz, $\omega_{b} = 2
  \pi \times 122$ Hz, $m_{a} = 1.44312 \times 10^{-25}$ Kg, $m_{b} = 1.40995
  \times 10^{-25}$ Kg, $a = 213 a_{0}$, $a_{12} = 99 a_{0}$ ($a_{0} = 5.3 \times 10^{-11}$ m, is
  the Bohr radius).}
\label{soliton}
\end{center}
\end{figure}

The soliton satisfies the boundary condition that, asymptotically as
$\sigma \rightarrow \sigma_{0}$, the phase variation vanishes. Here,
$\sigma_{0} = \frac{\mu}{g}$, is the equilibrium density of the
condensate. $u$ is the velocity of the soliton. The real part of Eq.
(\ref{GP1d_a}) can then be cast in a convenient form, in terms of
the densities:
\begin{eqnarray}
\sigma_{a} \sigma''_{a} - \frac{1}{2}\sigma'^{2}_{a} +
\left(\frac{1}{2}m_{a} u^{2} + \mu\right) \sigma^{2}_{a} &-& g
\sigma^{3}_{a} - \kappa \sigma_{b}\sigma^{2}_{a} \nonumber \\ &+&
\frac{1}{2} m_{a} u^{2}\sigma^{2}_{0} = 0. \label{density}
\end{eqnarray}

The following densities exactly solve Eq. (\ref{density}),
\begin{eqnarray}
\sigma_{a} &=& \sigma_{0} - \sigma_{0} \cos^{2}\theta
\textrm{sech}^{2}\left[\frac{\cos \theta}{\zeta} (x - u t)
  \right] \label{greysoliton}, \\ \textrm{and\,\,\,}
\sigma_{b} &=& b^{2} \textrm{sech}^{2}\left[\frac{\cos \theta}{\zeta} (x - u
  t) \right] \label{impurity},
\end{eqnarray}

\begin{figure}[t]
\begin{center}
\includegraphics[scale = 0.3]{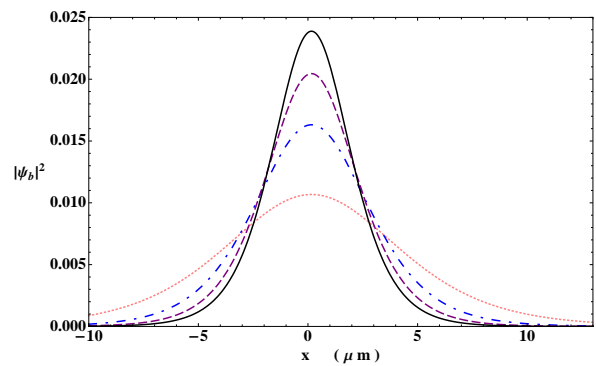}
\caption{The figure shows the variation of the density of the
impurity ($|\psi_{b}|^{2} = \sigma_{b}$) for different values of
atom-impurity coupling $\kappa$. The ratio $m_{b}/m_{a}$ is kept
fixed at $0.997$ for all the values of $g/\kappa$. The dotted
(pink), dash-dotted (blue), dashed (purple) and solid (black) lines
correspond to $a_{12} = 50 a_{0}, 100 a_{0}, 150 a_{0}$ and
$200 a_{0}$, respectively. One can see that as $a_{12}$, i.e.,
$\kappa$ increases, the density of the impurity increases. The
parameter values used for this figure are same as those of the Fig.
(\ref{soliton}).} \label{fig.2}
\end{center}
\end{figure}

where, $b^{2} = \frac{\cos \theta}{2 \zeta}$ with $\sigma_{0} = \frac{1}{2 \zeta \cos \theta
(g/\kappa - m_{b}/m_{a})}$ and $\hbar \omega =
\frac{\hbar^{2}k^{2}}{2 m_{b}} - \frac{\hbar^{2}}{2 m_{b}
\zeta^{2}}\cos^{2}\theta + \kappa \sigma_{0}$. The impurity atom
rests on the minimum of the grey soliton. Notice that for
$\sigma_{0}$ to be a positive definite, $\left(\frac{g}{\kappa} -
\frac{m_{b}}{m_{a}}\right) > 0$.  Repulsive atom-atom interaction
leads to $\kappa < \frac{g m_{a}}{m_{b}}$, provided the
atom-impurity interaction is also repulsive. If either of these two
interactions is attractive, the localized solutions cease to exist.
The presence of localized solitons crucially depends on the
balancing between nonlinearity and dispersion effects. The fact that
the condensate has the self-interaction, apart from its interaction
with the impurity, whereas, the impurity is devoid of
self-interaction, leads to the dominant role of the condensate
profile over the defect. It can be shown that for a significant
increase in the value of $g/\kappa$, corresponding to very weak
impurity coupling, the density of the impurity is very small and has
a marginal effect on the grey soliton. From Eq. (\ref{impurity}),
the following two cases $(a)$ $u = 0$, i.e., static solution and
$(b)$ $u \neq 0$, needs to be treated separately. When $u = 0$, the
limit $\kappa \rightarrow 0$ leads to vanishing of the impurity
wavefunction and the grey soliton tends towards a dark soliton
$\psi_{a} = \sqrt{\sigma_{0}} \tanh(\frac{x}{\zeta})$. However, this
limiting case leads to a divergence problem in the healing length
$\zeta = \frac{\hbar}{\sqrt{m_{b} \kappa \sigma_{0}}}$, and hence,
is not admissible. For the second case, it is clear from Eq.
(\ref{impurity}), that for $u \neq 0$, the limit $\kappa \rightarrow
0$ is unphysical, since the density of the impurity becomes
imaginary. In this case, the solutions exist only when $\kappa \geq
\frac{u^{2} m^{2}_{a}}{m_{b} \sigma_{0}}$. It is worth mentioning
that it has been recently found that, if the mass of the impurity
atom is too heavy and strongly interacting, the impurity can break
the condensate into two parts, where the mean-field theory breaks
down \cite{Lal}.

Interestingly, the sound velocity in this case is found to be:
$c_{w} = \zeta \sigma_{0} g/\hbar$. The velocity angle, also known
as the Mach angle, is given by,
\begin{equation}
\theta = \sin^{-1}\frac{u}{u_{s}}\label{sound},
\end{equation}

\begin{figure}[t]
\begin{center}
\includegraphics[scale = 0.35]{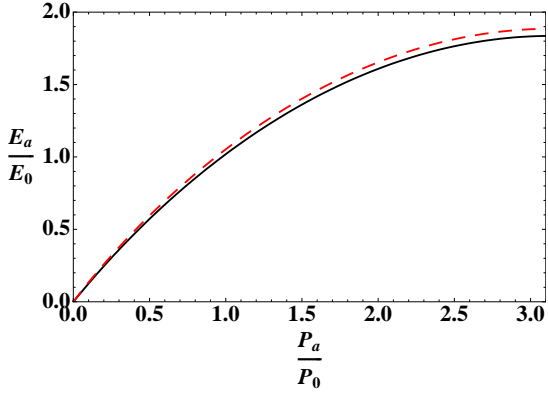}
\caption{Dispersion relation of the grey soliton (solid line), in
  presence of an impurity, with the dashed line above, depicting the
  same for pure grey soliton \cite{Jackson1}. The energy and momentum
  are measured in the units of $E_{0}$ and $P_{0}$, respectively. The
  parameter values, used for this figure are same as those of the Fig. (\ref{soliton}).}
\label{lieb}
\end{center}
\end{figure}

where the maximum velocity of the grey soliton is $u_{s} = c_{w}
\frac{m_{b} \kappa}{m_{a} g}$ for $\theta = \pi/2$. In the present
case, it is less than the sound velocity of the condensate. In the
limit $\kappa \rightarrow g m_{a}/ m_{b}$, though the soliton
maximum velocity $u_{s} \rightarrow c_{w}$, the equilibrium density
of the condensate $\sigma_{0}$ diverges. Therefore, the limit
$\kappa \rightarrow g m_{a}/ m_{b}$ is unphysical. The obtained
solutions exist within the limit $\frac{u^{2} m^{2}_{a}}{m_{b}
\sigma_{0}} \leq \kappa < g m_{a}/ m_{b}$. Superfluid density
vanishes at $\theta = 0$, which corresponds to a static soliton. For
a moving soliton, the superfluid density at the defect location is
finite.

In the presence of impurity, both the healing length and the maximum
soliton velocity get modified. The existence of the localized
soliton depends on the balance between nonlinearity and dispersion.
The atom-impurity nonlinear interaction contributes to the balancing
effect. As can be seen explicitly, the maximum velocity depends on
the density of the condensate, interaction between the atoms in the
condensate, as well as the interaction between the atoms and that of
the impurity. The presence of impurity atom lowers the grey
soliton's velocity. Interestingly, the effect of the impurity leads
to the appearance of the dimensionless ratio $\frac{m_{b}}{m_{a}}$,
in the maximum velocity expression.

The wavefunctions $\psi_{a}$ and $\psi_{b}$ can also be deduced:
\begin{eqnarray}
\psi_{a}(x,t) &=& i \sqrt{\sigma_{0}} \sin{\theta} + \sqrt{\sigma_{0}}
\cos{\theta} \tanh{\left(\frac{\cos{\theta}}{\zeta} (x - u
    t)\right)}, \label{wavefunc_greysol}\nonumber
\\ \\\psi_{b}(x,t) &=& i b
\textrm{\,sech}\left[(\frac{\cos{\theta}}{\zeta} (x - u t)\right) e^{i
  k x - i \omega t} \label{wavefunc_impurity}.
\end{eqnarray}

It is worth pointing out that the coupled complex envelope and bright
soliton solutions obtained here are of similar type, found earlier in
the context of two component BEC \cite{Busch} and in boson-fermion
mixtures \cite{Belmonte}. Fig. (\ref{soliton}) depicts the grey
soliton profiles, for various values of the Mach angle. Corresponding
defect profiles are shown in the inset. It is worth observing that the
localized defect resides at the minimum of the grey soliton, similar
to the experimental observation in a two component BEC, when one
component has less number of atoms \cite{Weiman}. Notice that, as
$\theta$ increases, the amplitudes of the grey soliton, as well as of
the defect decrease.

Fig. (\ref{fig.2}) shows the variation of the impurity density for
different values of $a_{12}$. The dotted (pink), dash-dotted (blue),
dashed (purple) and solid (black) lines correspond to $a_{12} =
50 a_{0}, 100 a_{0}, 150 a_{0}$ and $200 a_{0}$, respectively. One
can see that as $a_{12}$, i.e., $\kappa$ increases, the density of
the impurity increases. For small values of $\kappa$, the impurity
density is very small, which implies that the effect of the impurity
on grey soliton is marginal.

The stability of the grey soliton is investigated using the well
known criteria of Vakhitov and Kolokolov \cite{Vakhitov,Weinstein}
for non-linear Schr\"odinger type equation. It is known from this
criterion that the solution is stable and unstable if
$\frac{\partial N}{\partial \mu} > 0$ and $\frac{\partial
N}{\partial \mu} < 0$, respectively. When $\frac{\partial
N}{\partial \mu} = 0$, the solutions are found to be marginally
stable. We obtained the exact expression of the number of atoms from
the normalization condition of the condensate wavefunction,
\begin{equation}
N = \int^{\infty}_{-\infty} |\psi_{a}|^{2} dx = \frac{2 \hbar}{m_{b}
\kappa} \sqrt{\frac{m_{b}\kappa \mu - u^{2} m^{2}_{a} g}{g}}.
\end{equation}
In order to obtain the stability condition, we calculate,
\begin{equation}
\frac{\partial N}{\partial \mu} = \frac{2 \hbar}{g}
\left(\frac{m_{b}\kappa \mu - u^{2} m^{2}_{a}
g}{g}\right)^{-\frac{1}{2}}.
\end{equation}
The restriction on $\kappa$ gives $\frac{\partial N}{\partial \mu} >
0$. Therefore, the obtained solution is found to be stable within
the allowed range of $\kappa$.

The energy and momentum of the solitary wave and defect can be
obtained by removing the background contribution, which
ensures the convergence of the integrals,
\begin{eqnarray}
\label{simple_energy}
E_{a} = \frac{\hbar^{2}}{2 m_{a}}\int \frac{\partial
  \psi^{*}_{a}}{\partial x}\frac{\partial \psi_{a}}{\partial x} dx &+&
\frac{1}{2}g \int (\sigma_{a} - \sigma_{0})^{2} dx \nonumber \\ &+& \kappa \int
(\sigma_{a} - \sigma_{0})\sigma_{b} dx. 
\end{eqnarray}
Using the density profiles, obtained in Eq. (\ref{wavefunc_greysol})
and (\ref{wavefunc_impurity}), Eq. (\ref{simple_energy}) yields,
\begin{eqnarray}
E_{a} = \frac{2 \hbar^{2} \sigma_{0}}{3 m_{a} \zeta}\cos^{3} \theta +
\frac{2}{3} g \sigma^{2}_{0} \zeta \cos^{3}\theta - \frac{2}{3} \kappa \sigma_{0}
\cos^{2}\theta \label{energy}.
\end{eqnarray}

The three terms in the energy expression respectively represent, the
kinetic, self-interaction and the interaction energy between the
soliton and defect. As is evident, the energy of the grey soliton gets
substantially modified because of the presence of the impurity
implying the possibility of controlling soliton dynamics through the
defect. The momentum of the grey soliton remains unchanged: $P_{a} =
\sigma_{0} \hbar \left(\pi \frac{u}{|u|} - 2 \theta - \sin 2 \theta
\right)$. The dispersion relation is shown in Fig. (\ref{lieb}), which
also shows the same for that of the pure grey soliton for
comparison. The solid line depicts the dispersion of the grey soliton
in presence of an impurity, whereas, the dotted line is for pure grey
soliton. Both the energy and momentum are measured in the unit of
$E_{0} = \hbar \omega_{a}\sqrt{\sigma_{0} a}\sigma_{0}a_{\perp}$ and
$P_{0} = \sigma_{0} \hbar$ \cite{Jackson1}. It is observed that for long wavelength
excitations, the Lieb mode associated with the BEC-defect complex does not differ significantly with that
of the Lieb mode in defect free case at low momenta, and hence with the Bogoliubov mode. 
Hence, employing a single defect, one can control the velocity
of the grey soliton, which is found to be less than that of the pure
grey soliton.

For the sake of completeness, we have computed the energy and momentum of the impurity. The energy of the impurity with respect to the background $\sigma_{0}$
\begin{eqnarray}
E_{b} &=& \frac{\hbar^{2} k^{2}}{2 m_{b}} - \frac{\hbar \cos^{2} \theta}{6 m_{b} \zeta^{2}} - \frac{2}{3} \kappa
  \sigma_{0}\cos^{2} \theta,
\end{eqnarray}
with the corresponding momentum: $P_{b} = \hbar k$.

The solutions of the coupled mean-field equations also admit dark and
bright solitons, when both masses are equal ($m_{a} = m_{b} = m$):

\begin{eqnarray}
\psi_{a}(x,t) &=& \sqrt{\sigma_{0}}  \tanh[\frac{1}{\zeta} (x - u t)] e^{i k x - i \omega t},\nonumber \\  \\
\textrm{and\,\,\,\,} \psi_{b}(x,t) &=& b
\textrm{\,\,sech}[\frac{1}{\zeta} (x - u t)] e^{i k x - i \omega t},
\end{eqnarray}

with $m u = \hbar k$ and the healing length $\zeta =
\frac{\hbar}{\sqrt{m \kappa \sigma_{0}}}$. One notes that unlike the
previous case, a kinematic phase is associated with both atoms and
the impurity. The amplitude of the impurity is given by, $b =
\frac{1}{\sqrt{2 \zeta}}$ with $ \sigma_{0} = \frac{1}{2 \zeta
(\frac{g}{\kappa} - 1)}$. As is evident, for the solutions to exist,
the atom-atom coupling must be greater than atom-impurity coupling.
Furthermore, the solutions exist only when both the interactions are
repulsive or attractive. As compared to the previous case, the
velocity of the dark soliton $u$ can take any finite value.

In summary, we have obtained exact solutions for the grey soliton
impurity complex in a cigar-shaped geometry. The impurity resides in
the local minimum of the grey soliton, where the superfluid density
is finite. The velocity of the grey soliton gets restricted, which
in turn, controls the high momentum component associated with the
impurity. We observed that the presence of the impurity leads to a
reduction of grey soliton velocity, which for the defect free
condensate can reach sound velocity. It is found that the impurity
modifies the energy of the solitary wave, opening up the possibility
to control the grey soliton dynamics through impurity. The obtained
solutions are found to be stable, within the allowed range of
$\kappa$, as per the VK criteria. Pure dark or bright soliton,
having a vanishing superfluid component at the defect location, are
identified, when defect and condensate atoms have identical masses.
Unlike the previous case, there is no restriction on the velocity of
the grey soliton. We hope that grey soliton defect complex and its
dynamics can be observed with present laboratory setup. The response 
of this complex to trap and scattering length variations
\cite{Atre,Ranjani,Utpal} is worth investigating, as well as the
study of this complex in an optical lattice \cite{Priyam1,Priyam2}.

\end{document}